\documentclass[conference]{IEEEtran}
\IEEEoverridecommandlockouts
\usepackage{cite}
\newcommand\tab[1][0.3cm]{\hspace*{#1}}
\usepackage{amsmath,amssymb,amsfonts}
\usepackage{subfigure}
\usepackage{algorithmic}
\usepackage{graphicx}
\usepackage{textcomp}
\usepackage[font=small]{caption}

\usepackage{xcolor}
\def\BibTeX{{\rm B\kern-.05em{\sc i\kern-.025em b}\kern-.08em
    T\kern-.1667em\lower.7ex\hbox{E}\kern-.125emX}}
\begin{document}

\title{Transformer-Based Microbubble Localization\\
}

\author{\IEEEauthorblockN{Sepideh K. Gharamaleki}
\IEEEauthorblockA{\textit{Department of ECE} \\
\textit{Concordia University}\\
Montreal, Canada \\
sepideh.khakzadgharamaleki@mail.concordia.ca}
\and
\IEEEauthorblockN{Brandon Helfield}
\IEEEauthorblockA{\textit{Department of Physics and Biology,} \\
\textit{Concordia University}\\
Montreal, Canada \\
brandon.helfield@concordia.ca}
\and
\IEEEauthorblockN{Hassan Rivaz}
\IEEEauthorblockA{\textit{Department of ECE} \\
\textit{Concordia University}\\
Montreal, Canada \\
hrivaz@ece.concordia.ca}

}

\maketitle

\begin{abstract}
Ultrasound Localization Microscopy (ULM) is an emerging technique that employs the localization of echogenic microbubbles (MBs) to finely sample and image the microcirculation beyond the diffraction limit of ultrasound imaging. Conventional MB localization methods are mainly based on considering a specific Point Spread Function (PSF) for MBs, which leads to loss of information caused by overlapping MBs, non-stationary PSFs, and harmonic MB echoes. Therefore, it is imperative to devise methods that can accurately localize MBs while being resilient to MB nonlinearities, and variations of MB concentrations that distort MB PSFs. This paper proposes a transformer-based MB localization approach to address this issue. We adopted DEtection TRansformer (DETR) \cite{DETR}, which is an end-to-end object recognition method that detects a unique bounding box for each of the detected objects using set-based Hungarian loss and bipartite matching. To the authors' knowledge, this is the first time transformers have been used for MB localization. To appraise the proposed strategy, the pre-trained DETR network's performance has been tested for detecting MBs using transfer learning principles. We have fine-tuned the network on a subset of randomly selected frames of the dataset provided by the IEEE IUS Ultra-SR challenge organizers and then tested on the rest using cross-validation. 
For the simulation dataset, the paper supports the deployment of transformer-based solutions for MB localization at high accuracy.

\end{abstract}


\section{Introduction}
Recent advances in surpassing the resolution-depth limit of ultrasound imaging, inspired by Photo-Activated Localization Microscopy (PALM) \cite{PALM1}\cite{PALM2}, have led to microvessel angioarchitecture imaging via Ultrasound Localization Microscopy (ULM) \cite{ErricoULM}. ULM is a novel method for super-resolution imaging based on the localization and tracking of gas microbubbles (MBs). MBs, being approximately the size of a red blood cell, are contrast agents with strong acoustic backscattering that allows for the observation of small vessels with higher precision than conventional ultrasound. The increase in the intensity within blood vessels is due to the resonance of oscillating MBs and the impedance mismatch between blood and gas \cite{BHpaper}. \\
\tab Although ULM overcomes the diffraction limit of conventional ultrasound, it introduces another trade-off between MB localization accuracy and image acquisition time. The trade-off stems from the interference of echoes of multiple MBs that make the centroid of each MB indistinguishable from the other. Conversely, super-resolution ultrasound is achievable when the contrast agents are sufficiently isolated, i.e., separated by a wavelength or more. In light of this, diluted concentrations of MBs are utilized for localization, leading to long acquisition times. The prolonged data collecting procedure can prevent the detection of hemodynamic changes due to the unwanted effects of motion during the scan times. In \cite{95} and \cite{96}, low concentration of MBs were considered to avoid the issue of MB signal overlapping, sacrificing acquisition times.\\
\tab Numerous methods have investigated the low localization precision caused by MB signal overlapping and high MB concentration \cite{p1}-\cite{p3}. In \cite{p4}, a compressed sensing framework was deployed to reconstruct the images with high concentrations of MBs. Authors in \cite{p5} applied deconvolution to segregate single events of individual MBs in dense MB concentrations. These methods, while effective, render tracking individual MBs not plausible.\\
\tab Fourier-based filtering was also developed for the problem of MB signal separation \cite{p6} in high MB concentrations. Though it enhanced the ULM imaging quality, it was not as effective in higher MB counts or more complex flow hemodynamics and increased the computational cost of ULM.\\
\tab Deep learning (DL) algorithms using a variety of architectures have been broadly investigated in ULM imaging. DL-based methods were utilized to pinpoint MB centroids from B-mode images in the spatial domain with less computational complexity while maintaining high MB concentrations \cite{p7}. In \cite{p9}, a Convolutional Neural Network (CNN) was used to localize the MBs with RF channel data as the input. To resolve MB trajectories, the authors in \cite{p11} used a three-dimensional CNN as well as spatiotemporal information from the beamformed ultrasound dataset; all the while overriding ULM's conventional localization and tracking workflow. However, as a result, they were not able to produce velocity maps.\\
\tab Deep Super-resolution Micro-vessel Velocimetry (Deep-SMV) \cite{p12} is another localization-free method based on spatiotemporal data, which also provides velocity maps. Deep-SMV utilizes Long Short-Term Memory (LSTM) blocks which is a type of recurrent neural network (RNN). LSTMs, while more suited for learning temporal information, can be difficult to train owing to their long gradient paths.\\
\tab The utilization of Point Spread Functions (PSFs) and the localization of the centroid of each individual MB  poses yet another challenge in MB localization. Uncertainties in PSF estimation subject to variations in space, phase aberration, and attenuation lead to less reliable MB localization. Non-linear MB response, caused by generating harmonics, has proven to lead to further vulnerability in localization based on the centroids of the MB signals \cite{p13}.\\
\tab The MB localization error propagates along the ULM processing workflow and subsequently affects the ULM image quality, rendering localization a key concept in ULM imaging. Therefore, devising new methods for MB localization is an essential step toward more robust ULM imaging. Despite promising results in MB localization, a faster and better localization method providing high accuracy in different imaging settings remains to be developed. \\
\tab Considering the paradigm shift in natural language processing and image processing with the introduction of attention modules and transformers \cite{attention}, we believe it is time to focus on transformers for MB localization in ULM framework. An important advantage of transformers compared to CNNs is that they are not limited by the inductive bias of CNNs and can have substantially larger receptive fields with a self-attention module. Transformers also solve the issues of complex and extensive training of RNNs by parallelizing the operations which needed to be computed in serial before. To the best of our knowledge, Transformers in ultrasound were first used for classification of breast lesions~\cite{Gheflati _EMBC2022}. Transfer learning, which plays a significant role in medical imaging is also possible using Transformers.\\
\tab In this study, we focus on investigating the potential of Transformer-based solutions for MB localization for the first time. Our objective is to achieve comparable results to other DL-based algorithms while utilizing transfer learning. The advantage of using transfer learning is to avoid the need for large expert-annotated datasets. We test the proposed network on datasets collected with different imaging simulation configurations to manifest its capabilities. The data used in this study is released by the IEEE IUS 2022 challenge committee, Ultrasound Localisation and TRacking Algorithms for Super Resolution (Ultra-SR).

\section{Methods}

To study the efficiency of transformers in MB localization, we have adopted the DEtection TRansformer (DETR) network. The dataset provided by the challenge consists of two sets of ultrasound simulation videos with a high and low transmit frequency,  with and without ground truth of MB locations and tracking in all of the frames. \\
\subsection{COCO}
\tab Carrion \textit{et al.} \cite{DETR}, utilize COCO (Common Objects in COntext) format to mark the annotations of objects in each image as the input to the DETR to train and evaluate their proposed network. 
In order to form the annotations of MBs in ultrasound images from the IUS challenge dataset, a mask has been created on each frame of the video, presenting all the locations with MBs while preserving each of their information as an individual. Thereafter, the annotations of all MBs are individually found and stored in the COCO format.   \\ 
\subsection{DETR}
\begin{figure}[htb]
\begin{minipage}[b]{1\linewidth}
\centering
\includegraphics[width=\textwidth]{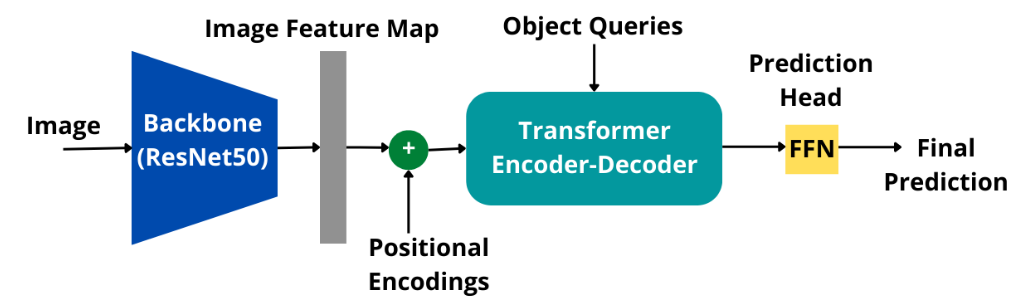}
\caption{DETR's architecture}
\label{DETR}
\end{minipage}
\end{figure}

The DETR consists of an encoder-decoder structure, with multi-head attention and self-attention as introduced in \cite{attention}. Attention allows the model to utilize the entire image as the context while deciding on all of the objects globally, considering their pair-wise correlations. \\
\tab An outline of the DETR's architecture is presented in Fig.~\ref{DETR}. The backbone to extract the image features is a CNN, in our case, ResNet50 \cite{resnet}. As part of the working mechanism, the transformer inputs the feature map pixels as query and key of its encoder-decoder structure.  The output of the decoder is mapped to a bounding box and class prediction using a 3-layer perceptron. There are in total 41M trainable parameters. \\
\tab Positional embeddings of the feature map is used for image features in the encoders and object features in the decoders in order to provide permutation-invariance in predictions. The object queries are learnt positional encodings that are added to the input of each attention layer in the decoder, as explained in the original paper. Since DETR uses set prediction loss, the number of object queries needs to be significantly larger than the typical number of objects in all the images. \\

\tab To avoid near-duplicates, DETR uses the direct set prediction loss, where pairs of predictions and targets are bipartite matched to assure unique bounding boxes for each object. To this end the ground truth has to be padded with “no object” classes, presenting the background of the images. 
The process of finding optimal matches, i.e. bipartite matching, is by minimizing the following loss function \cite{DETR}, 
\begin{equation}
\hat{\beta}=\arg \min \sum_{n=1}^{N} \mathcal{L}_{\operatorname{match}}\left(t_{n}, \hat{t}_{\beta(n)}\right)
\end{equation}
\normalsize
 where $\mathcal{L}_{\operatorname{match}}$ is defined as a linear combination of the class and bounding box error with ${t}_{n}=({c}_{n},{b}_{n})$ and $\hat{t} = ({\hat{c}}_{n},{\hat{b}}_{n})$ presenting the bounding box and the class associated with the ground truth and the predictions, respectively. \\
 
\tab The loss function is defined for the matches found via bipartite matching, i.e. $\hat{\beta}$, using the Hungarian algorithm, as follows \cite{DETR}:
\small
\begin{equation}
\mathcal{L}_{\text {Hungarian }}(t, \hat{t})=\sum_{n=1}^{N}\left[-\log \hat{p}_{\hat{\beta}(n)}\left(c_{n}\right)+1_{\left\{c_{n} \neq \emptyset\right\}} \mathcal{L}_{\text {box }}\left(b_{n}, \hat{b}_{\hat{\beta}(n)}\right)\right]
\end{equation}
\normalsize

with $\hat{p}_{\hat{\beta}(n)}\left(c_{n}\right)$ being the probability of class $c_{n}$ and $\mathcal{L}_{\operatorname{box}}$ representing a linear composition of the $\mathcal{L}_{\operatorname{1}}$ loss and the $\mathcal{G}_{\operatorname{IoU}}$ loss \cite{maybe giou} of the ground truth and the predicted bounding boxes.

\subsection{Transfer Learning}
Transfer learning has been integral in adopting DL methods in medical imaging. New tasks can leverage prior knowledge by developing on top of similar tasks learned in advance. The application of transfer learning has greatly benefited medical image analysis by reducing the computational complexity and the need for large annotated datasets. However, when it comes to natural vs medical images, there are fundamental dissimilarities, such as the data size, features, and the nature of tasks at hand. To adapt to the large domain shift between the two areas, we have fine-tuned the model on the challenge's dataset, using the pre-trained weights (trained on the COCO dataset \cite{coco}) of DETR. \\
\tab The learning rates are set the same as the DETR paper, and AdamW \cite{AdamW} is deployed for training. 70\%  of the data is used for training, and the rest for test using cross-validation.

\section{Results}
\subsection{Simulation dataset with ground truth}
\begin{figure}
  \centering
  \subfigure{\includegraphics[scale=0.5]{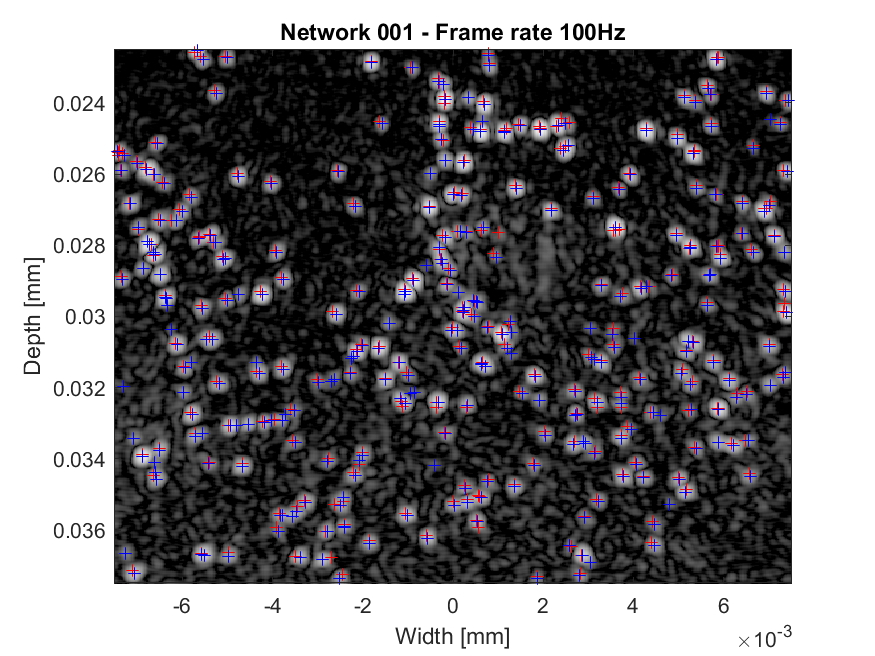}}
  \caption*{(a) Linear array}\quad
  \subfigure{\includegraphics[scale=0.5]{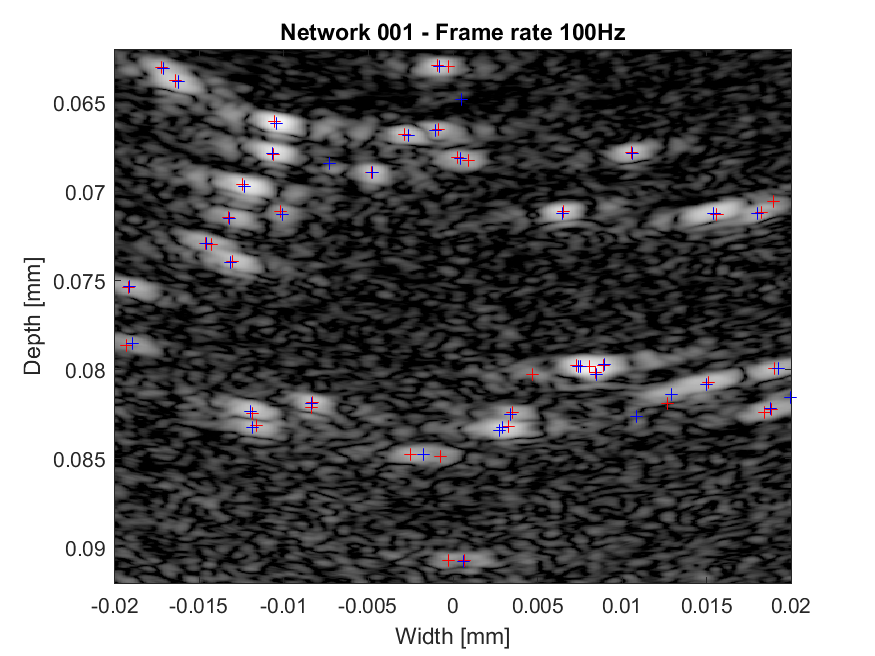}}
  \caption*{(b) Phased array}
  \captionsetup{justification=centering}
  \caption{Results of localization on the high frequency  data with ground truth (Red: predictions, Blue: ground truth)}
  \label{wgt}
\end{figure}

The performance of the proposed network was quantified by fine-tuning on the IUS challenge dataset. The focus of the evaluation is the ability to accurately identify individual MB locations. Since, as previously mentioned, the number of object queries needs to be significantly higher than the maximum number of objects in all the images, we have divided each frame of the videos provided by the challenge into four windows, to avoid increasing the object queries.\\
\tab The experiments were executed on a workstation running Ubuntu 20.04 operating system, with Intel® Core™ i7-11700F @ 2.50GHz CPU (16 cores). The fine-tuning takes 1 hour and 28 minutes using NVIDIA® GeForce® RTX 3080 Ti GPU and 64GB of RAM.\\
\tab Fig.~\ref{wgt} shows the final output for one frame of each video with ground truth. To evaluate the localization results, we have used mean Average Precision (mAP) and mean Average Recall (mAR) over all IoU thresholds. The final mAP and mAR are 83.4\% and 55.2\%, respectively.

\subsection{Simulation dataset without ground truth}
\begin{figure}
  \centering
  \subfigure{\includegraphics[scale=0.5]{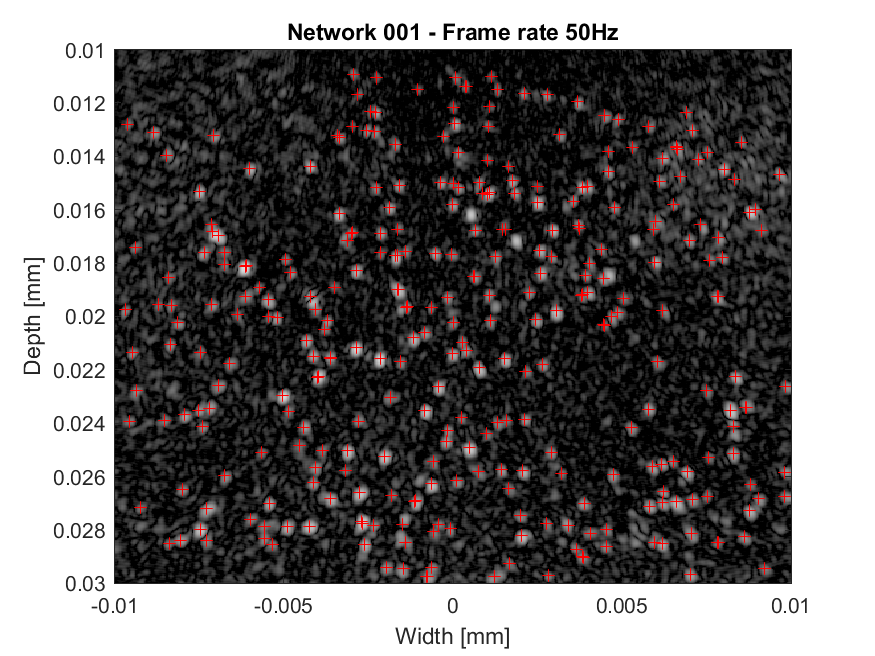}}
  \caption*{(a) Linear array}\quad
  \subfigure{\includegraphics[scale=0.5]{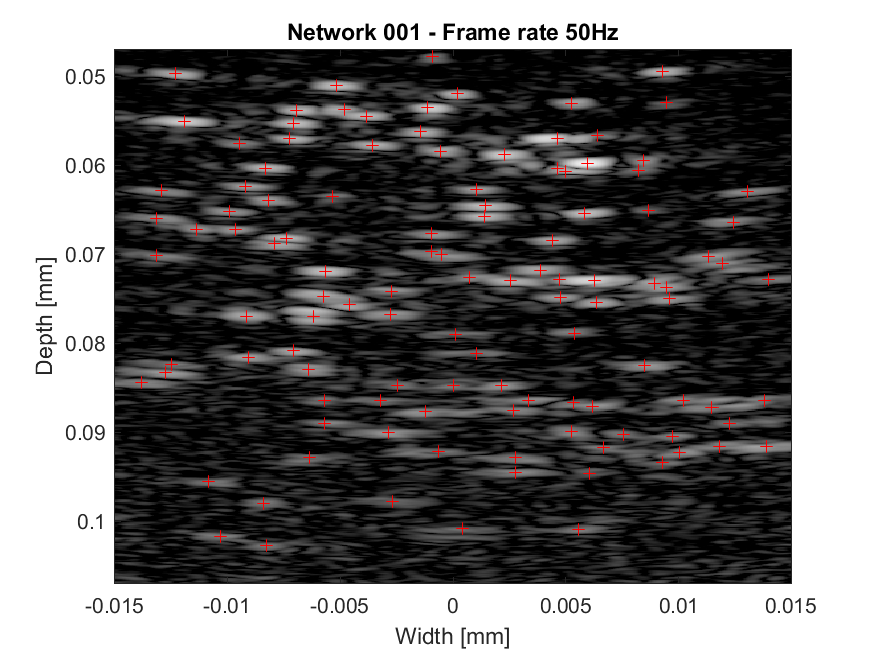}}
  \caption*{(b) Phased array}
  \captionsetup{justification=centering}
  \caption{Results of localization on the low frequency data without ground truth (Red crosses: predictions)}
  \label{w/gt}
\end{figure}

Trial-and-error for the dataset without ground truth manifested better results after patching the frames into nine windows. An algorithm was also devised to avoid duplicate localizations for MBs on the borders of the patches. Furthermore, to account for different imaging configurations, such as phased or linear arrays or different frequencies, we have fine-tuned the network on the first frame of each video. Fig.~\ref{w/gt} shows the network's results on the dataset without ground truth.

\section{Conclusions}

In this study, we investigated the performance of a transformer-based approach for MB localization in ULM under different imaging configurations. We successfully assessed the potential of DETR for MB localization, while also avoiding pre-processing the MB PSFs. 
By adopting transfer learning, we also circumvented time and hardware-consuming computations. We found that fine-tuning the network weights on a frame of the acquired images enhances MB localization results by reducing the number of false negatives. This is necessary to make the algorithm compatible with a variety of imaging configurations. We have utilized patching as a remedy for DETR’s low intrinsic performance on small objects and developed an algorithm to handle the MBs that fall on the border of these patches in order to avoid duplicate MB centers.  \\
DETR's long training time ($>$300 epochs) could be a limitation of this study. Future studies can focus on other Transformer-based approaches such as deformable-DETR \cite{deformabledetr} to address this issue. 
The results of this work should also be tested for the in-vivo data to evaluate its real-world applications.
Other ULM processing steps such as MB tracking has not been covered in this study. Considering transformers' highly efficient tracking capabilities, pursuing future research into tracking MBs while employing them could be a fascinating next step.
Finally, more recent Transformer-based networks for detection and localization that perform better for small objects can be utilized to improve the localization accuracy.

\section*{Acknowledgment}
We acknowledge the support of the Natural Sciences and Engineering Research Council of Canada (NSERC). B.H. holds a Burroughs Wellcome Fund CASI award

\end{document}